\documentclass[conference]{IEEEtran}
\IEEEoverridecommandlockouts
\usepackage{cite}
\usepackage{amsmath,amssymb,amsfonts}
\usepackage{algorithmic}
\usepackage{graphicx}
\usepackage{epstopdf}
\usepackage{amsthm}
\newtheorem*{remark}{Remark}
\usepackage{multirow}
\graphicspath{{Images/}}
\ifCLASSOPTIONcompsoc
\usepackage[caption=false,font=normalsize,labelfont=sf,textfont=sf]{subfig}
\else
\usepackage[caption=false,font=footnotesize]{subfig}
\fi
\usepackage{stfloats}
\usepackage{textcomp}
\usepackage{xcolor}
\def\BibTeX{{\rm B\kern-.05em{\sc i\kern-.025em b}\kern-.08em
    T\kern-.1667em\lower.7ex\hbox{E}\kern-.125emX}}
\begin{document}

\title{Robust and Optimal Single-loop Voltage Controller for Grid-forming Voltage Source Inverters
}

\author{\IEEEauthorblockN{Soham Chakraborty,
Sourav Patel, and Murti V. Salapaka}
\IEEEauthorblockA{Department of Electrical and Computer Engineering\\
University of Minnesota
\\ \{chakr138, patel292, murtis\}@umn.edu}}


\maketitle

\begin{abstract}
Design and implementation of an optimal and robust single-loop voltage controller is proposed for single-phase grid-forming voltage source inverter (VSI). The objective of the proposed controller is to have good reference tracking and disturbance rejection. Uncertain nature of loads can significantly alter system's behavior, especially during heavily loaded conditions, and impose an uncertainty in the dynamic model. This deteriorates the robustness of the controller severely. $\mathcal{H}_{\infty}$-based controller design is proposed in this article to address this issue. The required objectives of the optimal controller are formulated after modeling the loads as uncertain element of the control system. Time-domain MATLAB/SIMULINK-based simulation study substantiates the fact that the resulting controller exhibits superior robustness in performance during varying loading conditions of VSI than that of the conventional multi-loop control architecture. OPAL-RT based controller hardware-in-the-loop (CHIL) simulations with low-cost controller are conducted to validate the computational footprint of the resulting controller.
\end{abstract}
\section{Introduction}
During islanded mode of operation of AC microgrid, grid-forming voltage source inverters (VSI) play an essential role by maintaining a stable voltage and frequency of the network and supplying uninterrupted power to the loads in the absence of a stiff distribution grid. Both centralized (\textit{master-slave}) and decentralized (\textit{droop-based}) \textit{Level-1} control layer architecture have been proposed in order to meet the uninterrupted power supply to the loads in the network via generating a voltage reference for the \textit{Level-0} control layer \cite{masterslave}. In \textit{Level-0} control layer, a voltage controller is required to be designed for each grid-forming VSI so that the output voltage waveform of each VSI tracks the respective voltage reference signal \cite{microgrid}.
\par Designing the voltage controller is essentially a multi-objective task. The design needs to primarily guarantee perfect reference tracking and good disturbance rejection which results in an output voltage with good regulation and low harmonic distortion in presence of various linear and non-linear loads. Additionally, the controllers are required to provide compensation to dynamic variations of the output current and improve the dynamic response \cite{yazdani}. It is also found that the unknown nature of the output loads can significantly alter the system behavior, especially during heavily loaded condition of VSI, that deteriorates the transient performance severely \cite{loaduncertainty}. The controller is required to perform all the aforementioned multi-objective tasks under output load uncertainties. As a result, requirement of robust performance of the voltage controller warrants an essential criterion during design stage. 
\par Numerous voltage control strategies are proposed in the literature during past decades for grid-forming VSIs. Most of the earlier control techniques are based on linear control theory with a single voltage control-loop scheme \cite{guerrero1}. To further enhance the disturbance imposed by variations in input voltage and output current, a non-linear feed-forward loop is added in \cite{guerrero2}. However, single-loop classical controllers experience poor transient response and limited stability margins \cite{multiloop1}. To overcome these drawbacks, multiple-loop proportional-resonant (PR) and proportional-integral-differential (PID)-based classical control methods are proposed in \cite{multiloop1,multiloop2,multiloop3,multiloop4}. This common architecture has an internal current negative feedback loop, commanded by the error signal of the outer voltage regulation loop \cite{yazdani}. By introducing the inner current loop, the poles of the resonant $LC$ filter can be decoupled to compensate the non-idealities of the inductor. Moreover, it facilitates an inherent over-current protection of the VSI. However, having an inner-loop current controller requires more sensors that increases the overall cost \cite{prodanovic}. Moreover, classical controllers lack robustness in performance with output load uncertainties \cite{multiloop3}. In \cite{zhong1,zhong2,zhong3}, a lower order controller is designed using $\mathcal{H}_{\infty}$ methodology by removing the higher order part of the internal model using repetitive control technique. However, in these approaches load uncertainty is not taken into account. A robust tuning strategy for PID-based double-loop lag–lead compensator for VSI is proposed in \cite{multihinf1}. $\mathcal{H}_{\infty}$-based methodology is further explored in \cite{multihinf2,multihinf3,multihinf4,hinf2} in order to design an optimal controller for VSIs with linear and non-linear load disturbances. However, both these works inherently adapt the multi-loop structure for voltage controller and load uncertainty is not considered during design. In \cite{h2hinf1}, a sliding-mode control-based design for inner current loop and a mixed $\mathcal{H}_2/\mathcal{H}_{\infty}$-based optimal design for outer voltage loop is proposed. A very similar design is proposed in \cite{h2hinf2} where linear matrix inequality (LMI)-based linear quadratic regulator (LQR) for inner current loop and mixed $\mathcal{H}_2/\mathcal{H}_{\infty}$-based design for outer voltage loop are employed. Both these designs improve the transient response significantly in presence of uncertain filter parameters at the cost of added complexity in controllers. Moreover, uncertainty imposed by output loads is not taken into consideration while designing the controllers. \cite{hinf1} provides a robust optimal state-feedback controller based on the integration of optimal output regulation theory and back-stepping method where load is treated as a uncertain element. However, the impact of non-linear loads (various harmonic components) is not considered in this work. The primary contributions of this article are as follows:
\begin{enumerate}
    \item A simple robust and optimal (single-loop, low-cost) voltage controller for \textit{Level-0} control layer is proposed that can address the aforementioned limitations along with taking into account the uncertainties in load variations.
    \item It facilitates lower requirements of sensor measurements as opposed to other works in the literature without compromising in voltage regulation and disturbance rejection while guaranteeing robust performance.
\end{enumerate}
Time-domain MATLAB/SIMULINK-based simulation with low-cost controller are performed in the validation stage of the proposed control.
\begin{figure*}[t]
	\centering
	\subfloat[]{\includegraphics[scale=0.28,trim={18cm 2cm 14cm 3cm},clip]{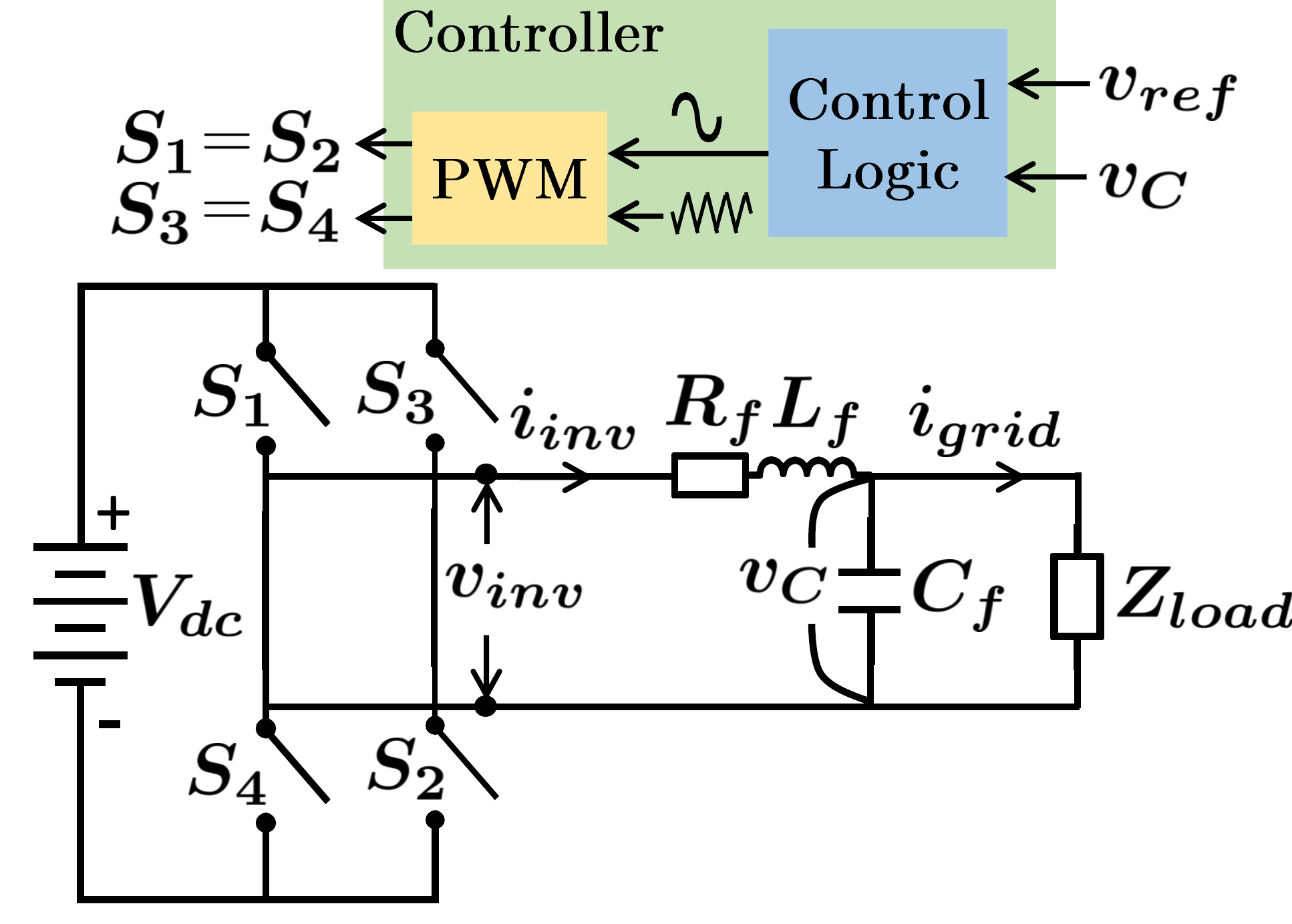}%
		\label{fig:circ}}~~~~~~~~
	\subfloat[]{\includegraphics[scale=0.18,trim={0cm 0cm 0.0cm 0.0cm},clip]{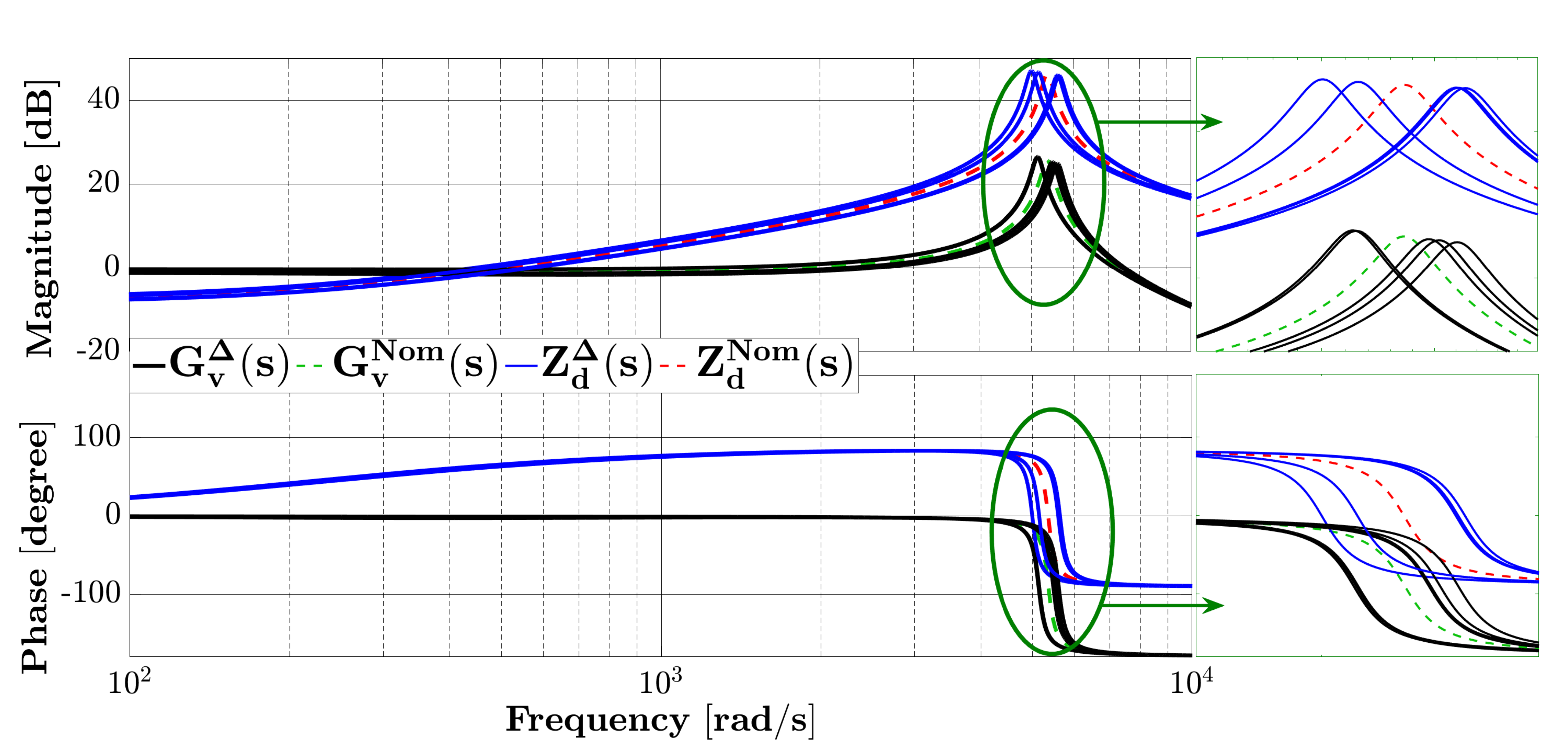}%
		\label{fig:GvZd}}
	\caption{Description of the system with (a) the schematic of single-phase grid-forming VSI, (b) variation of plant dynamics model due to uncertain load.}
	\label{fig:circuit}
\end{figure*}
\section{Problem Formulation}
\subsection{Architecture of Grid-forming VSI}
A single-phase grid-forming VSI terminated by a load, $Z_{Load}$, is shown in Fig.~\ref{fig:circuit}\subref{fig:circ}. The VSI is composed of a dc bus voltage, $V_{dc}$, four switching devices, $S_1$-$S_4$, and an $LC$ filter with $L_f$, $C_f$ as filter inductor and capacitor respectively. $R_f$ is the equivalent series resistance (ESR) of $L_f$ capturing parasitic element. The controller along with sinusoidal PWM switching technique generates the switching signals that result the output voltage. The dynamics of the VSI are described as:
\begin{align}
L_f \frac{d\langle i_{inv}\rangle}{dt} +R_f\langle i_{inv}\rangle &= \langle v_{inv}\rangle - \langle v_C\rangle, \label{eq1}\\
C_f \frac{d\langle v_C\rangle}{dt} &=  \langle i_{inv}\rangle - \langle i_{grid}\rangle, \label{eq2}
\end{align}
where $\langle .\rangle$ signifies the average values of the corresponding variable over one switching cycle ($T_s$) \cite{yazdani}. Combining \eqref{eq1} and \eqref{eq2} and taking Laplace transformation, the open-loop output voltage dynamics of the grid-forming VSI is given as:
\begin{align}\label{eq3}
V_C(s) &= G_{inv}(s)V_{inv}(s) - G_{i}(s)I_{grid}(s),
\end{align}
where 
\begin{align*}
    G_{inv}(s) &= \dfrac{1}{L_f C_f s^2 +R_f C_f s +1},\\ G_{i}(s) &= \dfrac{L_f s+R_f}{L_f C_f s^2 +R_f C_f s +1}.
\end{align*}
\subsection{Load Characteristics and Modeling}
Characterizing the nature of the load is essential in the process of designing the controller. In this work, the load is modeled by a parallel combination of two components. First component is an admittance, $Y_L(s)$, with unknown $R$ and $L$ elements, and can be defined as:
\begin{align*}
    Y_L(s):= [1+\Delta(s)]Y_L^N(s),
\end{align*}
where, $Y_L^N(s)$ is a frequency dependent weighting function, capturing the nominal behavior, and defined as:
\begin{align*}
    Y_L^N(s):=\dfrac{1}{L_L^Ns+R_L^N}.
\end{align*}
A normalized dynamic LTI uncertainty, $\Delta(s)$, is defined to capture the uncertain behavior of load change such that
\begin{align*}
    ||\Delta(s)||_{\infty} \leq 1.
\end{align*}
The value of $L_L^N$ and $R_L^N$ can be selected as $R_L^N = V_{rated}^2/P_{rated}$ and $L_L^N = V_{rated}^2/(\omega_oQ_{rated})$, where $V_{rated}$, $P_{rated}$, $Q_{rated}$ and $\omega_o$ are the rated output voltage, active, reactive power of the grid-forming VSI and nominal frequency of the network respectively. Another component of loads is a parallel combination of current sources consisting of both fundamental and harmonic components \cite{loadmodel1} and defined as
\begin{align*}
    I_d(s):=\sum_{h}I_{d,h}(s),~~ [~h~\text{is odd}~].
\end{align*}
Therefore, $Z_{Load}$ of Fig.~\ref{fig:circuit}\subref{fig:circ} can be characterized as
\begin{align}\label{eq4}
    I_{grid}(s) &= Y_L(s)V_C(s) + I_d(s) \nonumber \\
    &= [1+\Delta(s)]Y_L^N(s)V_C(s) + \sum_{h}I_{d,h}(s).
\end{align}
\begin{remark}
Multiplicative uncertainty in $Y_L(s)$ is used to shape the ``worst-case'' yet allowable output admittance of the grid-forming VSI. The exogenous disturbance signal, $I_d(s)$, captures the presence of non-linear loads as shown in Fig.~\ref{fig:control}\subref{control1}.
\end{remark}
\subsection{Impact of Load Uncertainty}
Combining \eqref{eq3} and \eqref{eq4}, system's dynamic equation becomes
\begin{align}\label{eq5}
    V_C(s) = G_v^{\Delta}(s)V_{inv}(s) - Z_d^{\Delta}(s)I_d(s).
\end{align}
where, 
\begin{align*}
    G_v^{\Delta}(s) = \dfrac{G_{inv}(s)}{1+G_i(s)Y_L(s)},~~ Z_d^{\Delta}(s) = \dfrac{G_i(s)}{1+G_i(s)Y_L(s)}.
\end{align*}
Clearly uncertainty in $Y_L(s)$ imposes uncertainty in both $G_v^{\Delta}(s)$ and $Z_d^{\Delta}(s)$ as illustrated in Fig.~\ref{fig:circuit}\subref{fig:GvZd}. It can be observed that the unknown load across VSI causes significant change in the plant dynamics, especially the effective resonant frequency of the system. This deteriorates the robust performance of the controller severely as well as overall system stability \cite{loaduncertainty}.
\section{Proposed Solution}
\subsection{Objectives of the Designed Controller}
The objective is to design a feedback control law through controller, $C_{H_{\infty}}(s)$ as shown in Fig.~\ref{fig:control}\subref{control1}, which generates a control signal, $v_{inv}$, such that, i) $v_C$ tracks $v_{ref}$, ii) effects of $i_d$ on $v_C$ is attenuated, iii) $v_{inv}$ satisfies bandwidth limitations, and iv) $v_C$ recovers quickly after any sort of transients. Moreover, all aforementioned objectives need to be achieved with load uncertainty. In other words, it is necessary to provide the controller with enough robustness to deal with system uncertainty caused by load variation. These objectives are derived from acceptable standards on microgrid operation, IEEE Std-2030 \cite{ieee1}, and power quality, IEEE Std-519\cite{ieee2}.
\subsection{Design Procedure of the $\mathcal{H}_{\infty}$-based Controller}
\begin{figure*}[t]
	\centering
	\subfloat[]{\includegraphics[scale=0.25,trim={5cm 1.0cm 8.5cm 2.5cm},clip]{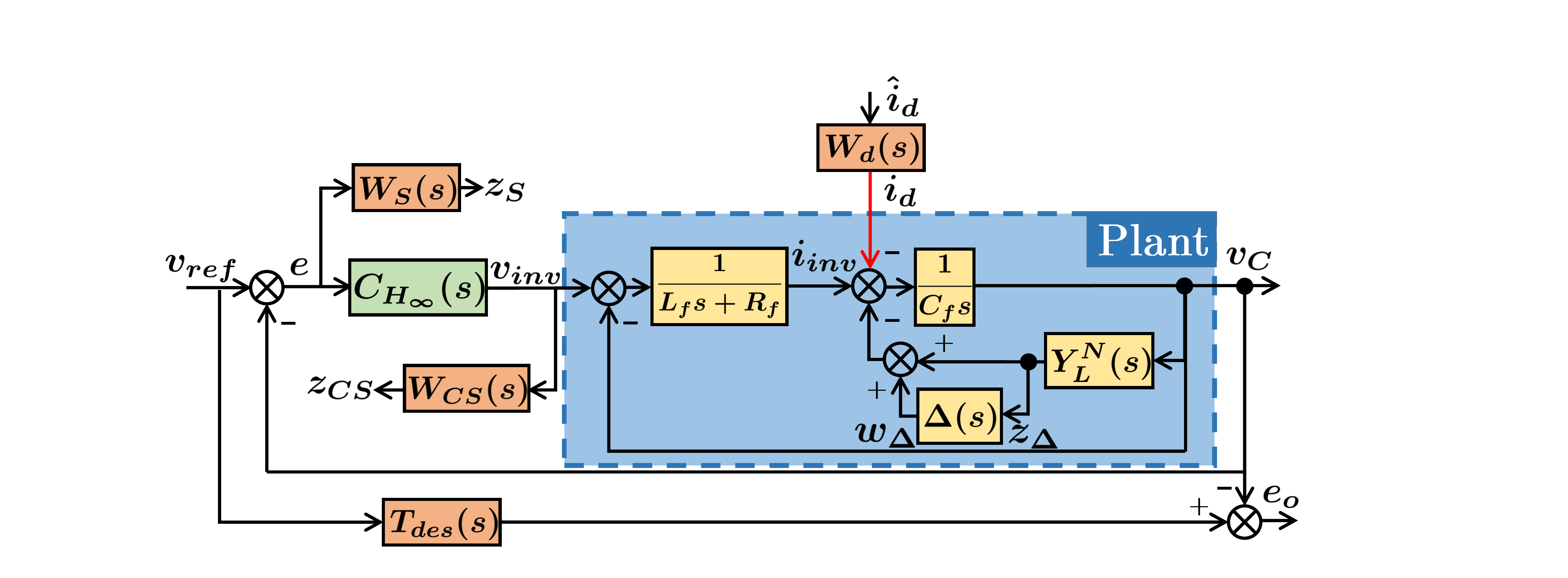}%
		\label{control1}}~~~~~~~~~
	\subfloat[]{\includegraphics[scale=0.27,trim={19.5cm 5.5cm 19.0cm 3.5cm},clip]{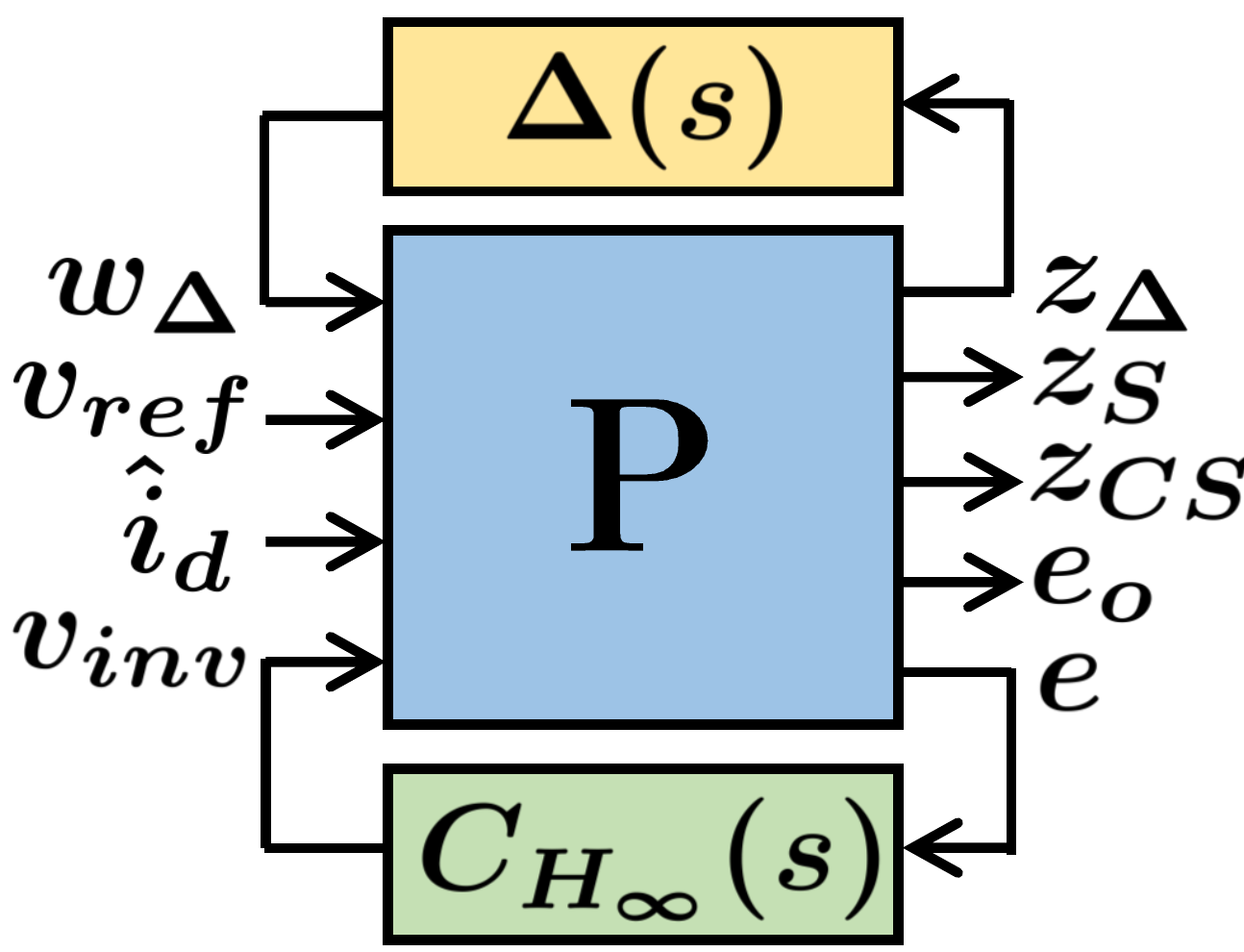}%
		\label{control2}}
	\caption{$\mathcal{H}_{\infty}$-based controller synthesis, (a) the closed-loop control system with selected weighting transfer functions, (b) general control configuration.}
	\label{fig:control}
\end{figure*}

$\mathcal{H}_{\infty}$-based controller design provides a framework for addressing multiple objectives. Here, the design is based on the system structure presented in Fig.~\ref{fig:control}\subref{control1} where user-defined weighting transfer functions, $W_S(s)$, $W_{CS}(s)$, $W_d(s)$ and $T_{des}(s)$, are selected based on the aforementioned requirements. The brief guidelines for designing the weighting transfer functions are provided below.
\subsubsection{Selection of $W_S(s)$}
To shape the sensitivity transfer function, the weighting function, $W_S(s)$ is introduced so that
\begin{itemize}
    \item the tracking error at $\omega_o$ is extremely low,
    \item the $LC$ resonance of VSI is actively damped,
\end{itemize}
$W_S(s)$ is modeled to have peaks at $\omega_o$ and $LC$ resonant frequency, $\omega_{res}$, with a overall gain $k_{S,1}$. Hence, 
\begin{align*}
    &W_S(s) \\
    & = k_{S,1}\bigg[\dfrac{s^2+2k_{S,2}\zeta\omega_o s+\omega_o^2}{s^2+2\zeta\omega_o s+\omega_o^2}\bigg] \bigg[\dfrac{s^2+2k_{S,3}\zeta\omega_{res}s+\omega_{res}^2}{s^2+2\zeta\omega_{res}s+\omega_{res}^2}\bigg],
\end{align*}
where, $k_{S,2}$ and $k_{S,3}$ are selected to exhibit peaks and $\zeta$ takes care of the off-nominal system frequency.
\subsubsection{Selection of $W_{CS}(s)$}
$W_{CS}(s)$ is designed to suppress high-frequency control effort to shape the performance of $v_{inv}$. Hence, it is designed as high-pass filter with cut-off frequency at switching frequency and ascribed the form:
\begin{align*}
    W_{CS}(s) = \dfrac{s+k_{CS,1}\omega_o}{s+k_{CS,2}\omega_o}, ~\text{where}~ k_{CS,1}<<k_{CS,2}.
\end{align*}
\subsubsection{Selection of $W_d(s)$}
$W_d(s)$ emphasizes the expected disturbances at $\omega_o$ and at different harmonic frequencies imposed by $i_d$ and emphasized by exogenous signal $\hat{i}_d$. It is based on the assumption that the load current comprises fundamental, $3^{rd}$, $5^{th}$, $7^{th}$, $9^{th}$, $11^{th}$ and $13^{th}$ harmonics. Hence, it is designed by a low-pass filter with peaks at selected frequencies. The resulting function is obtained as
\begin{align*}
    W_d(s) = \prod_{h=1,3,\ldots,13}\dfrac{s^2+2k_{d,h}\zeta h\omega_o s+h^2\omega_o^2}{s^2+2\zeta h\omega_o s+h^2\omega_o^2},
\end{align*}
where, the values of $k_{d,h}$ are selected based on the THD standards mentioned in IEEE Std-519 \cite{ieee2}. 
\subsubsection{Selection of $T_{des}(s)$}
To emphasize on the desired performance of the closed-loop system, a performance bound is designed with a typical first-order transfer function specified by the parameters $k_B$, $\omega_B^*$ in the following expression.
\begin{align*}
    T_{des}(s) = k_B\dfrac{\omega_B^*}{s + \omega_B^*}.
\end{align*}
Clearly, the desired output response is decided based on the selection of $\omega_B^*$, which is determined by the requirement of speed of response. Desired bandwidth of at least $10$ times of $\omega_o$, decides the value of $\omega_{B}^*$ with $k_B$ as overall gain. 
\begin{figure}[t]
	\centering
	\subfloat[]{\includegraphics[scale=0.25,trim={1.0cm 0.0cm 2.2cm 0cm},clip]{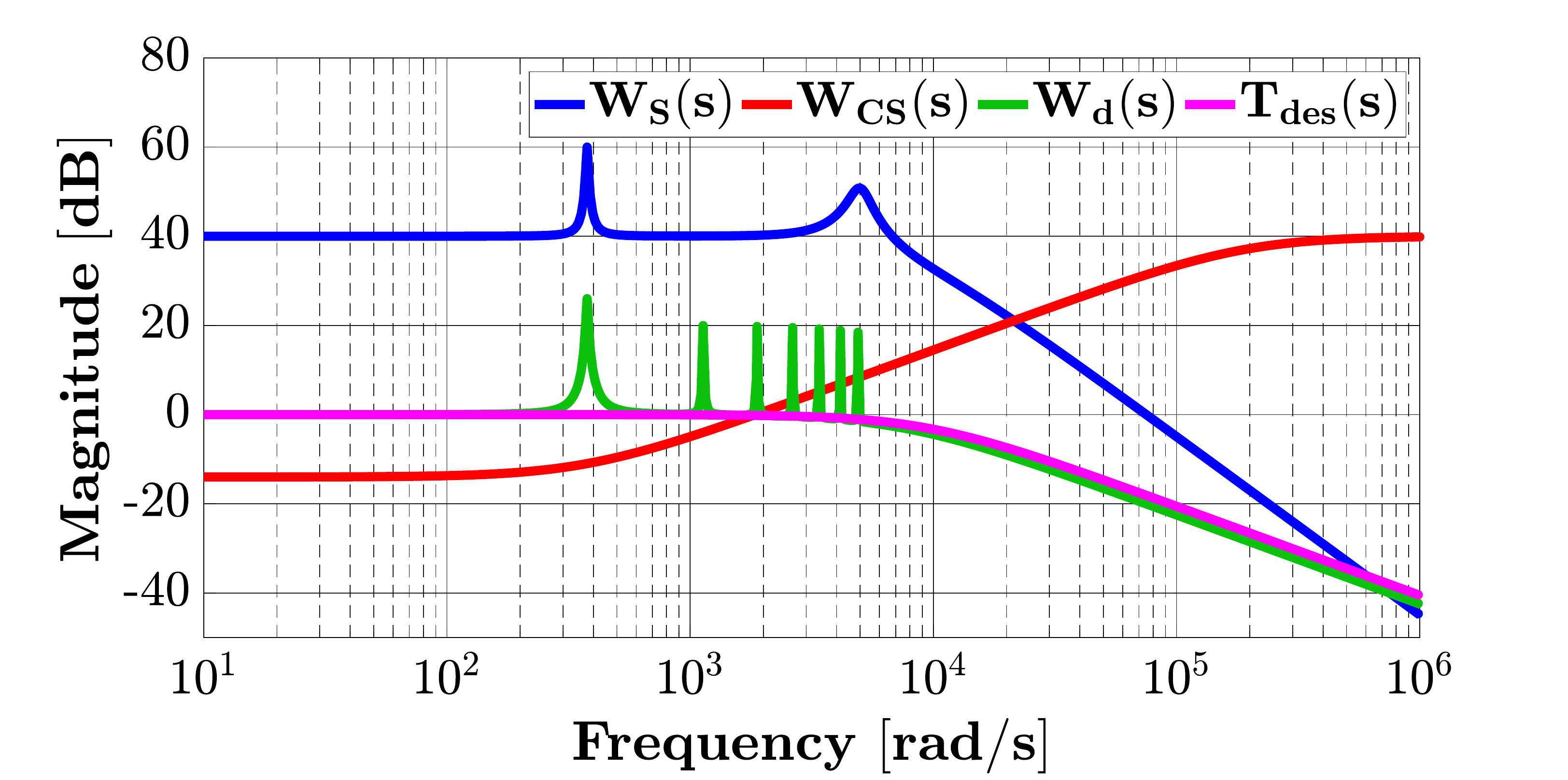}%
		\label{fig:weight}}
		
	\subfloat[]{\includegraphics[scale=0.25,trim={1.0cm 0.0cm 2.2cm 0cm},clip]{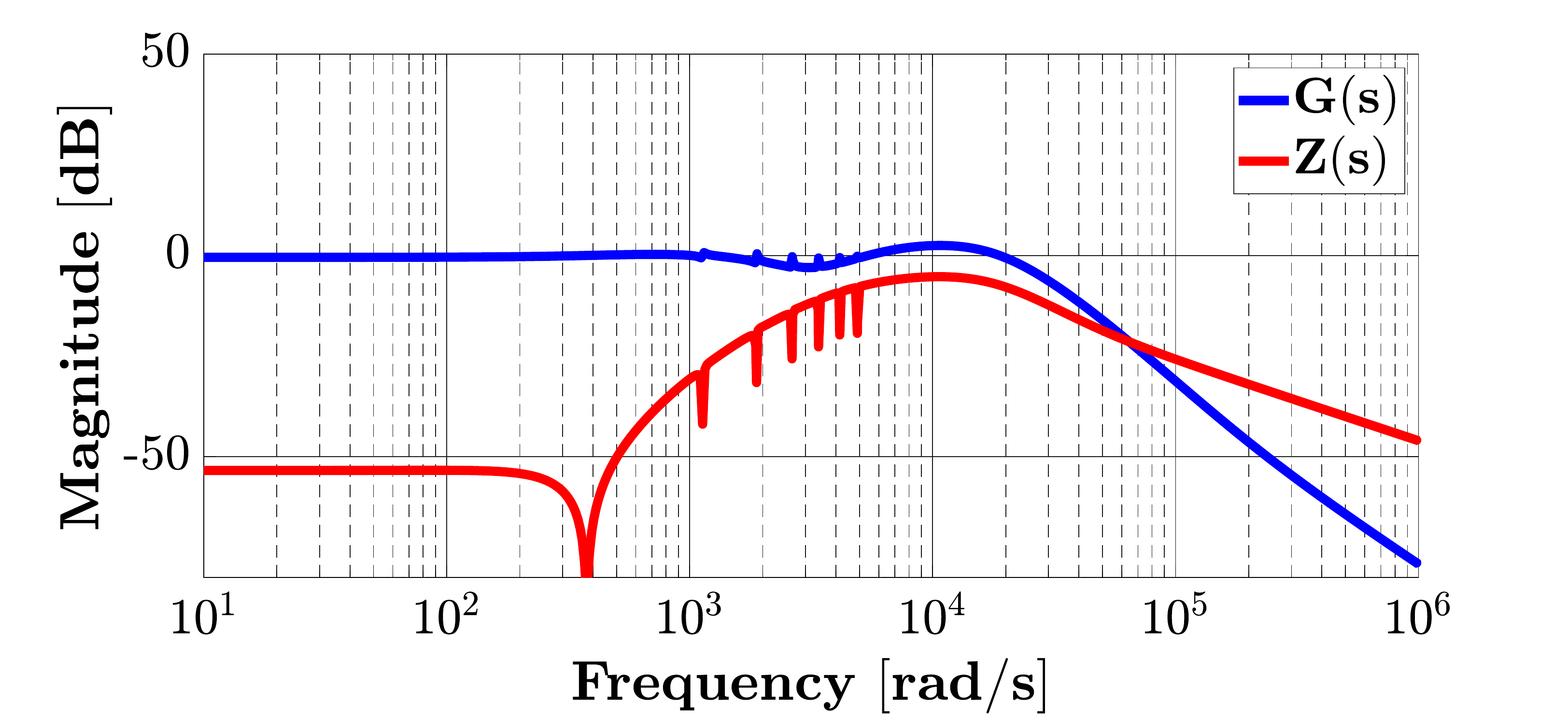}%
		\label{fig:GZmag}}
		
    \subfloat[]{\includegraphics[scale=0.25,trim={0.5cm 0.0cm 2.2cm 0.0cm},clip]{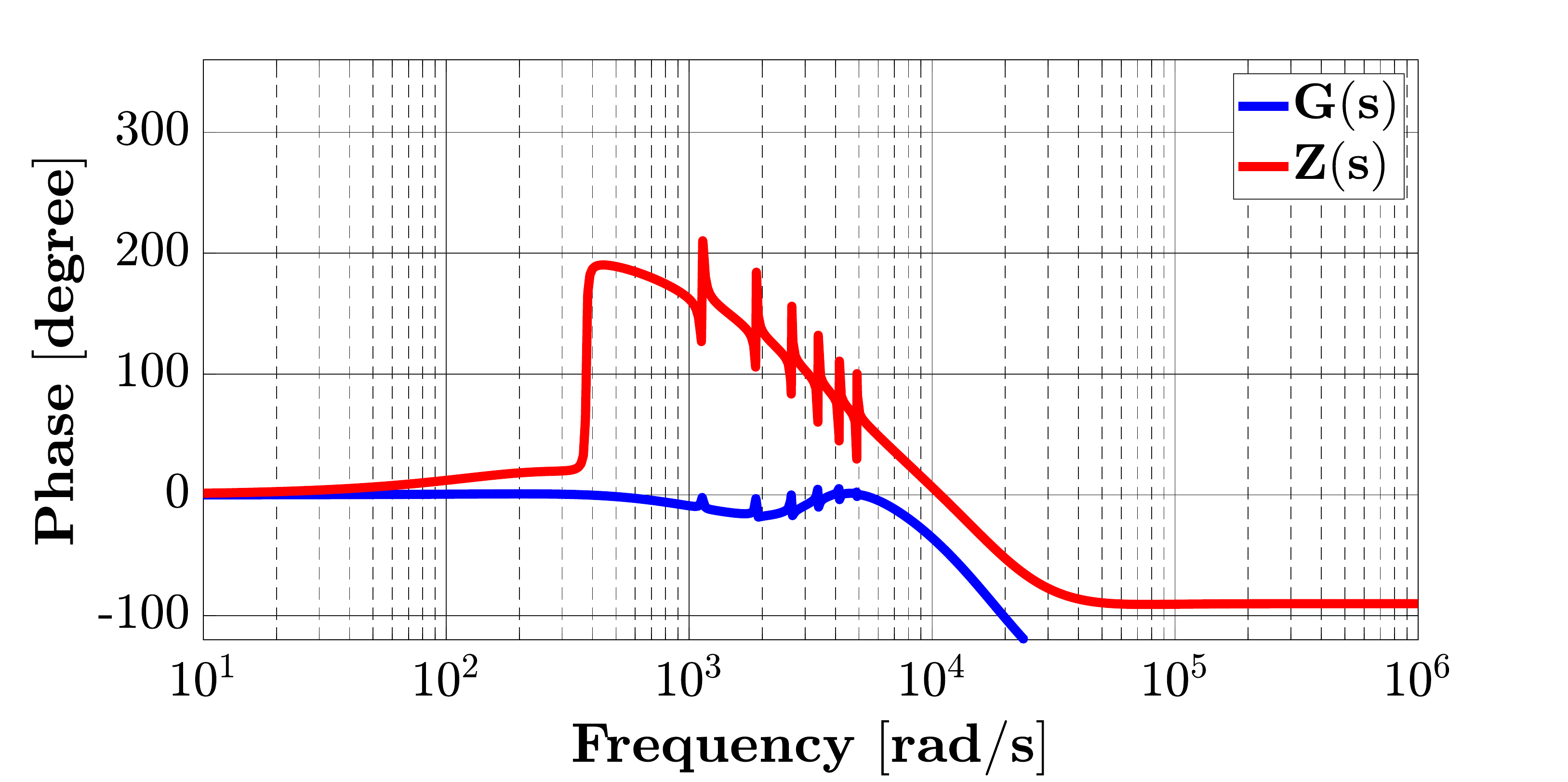}%
		\label{fig:GZphase}}
	\caption{Bode plots of (a) magnitudes of weighting transfer functions, (b) magnitudes of $G(s)$ and $Z(s)$ and (c) phase of $G(s)$ and $Z(s)$.}
	\label{fig:bode}
\end{figure}
\subsection{Problem Formulation and Resulting Optimal Controller}
The Bode plots of the selected weighting transfer functions in this work are shown in Fig.~\ref{fig:bode}\subref{fig:weight}. The $\mathcal{H}_{\infty}$ optimal problem is formulated to generate a feedback control law with controller, $C_{H_{\infty}}(s)$, stated as:
\begin{align}\label{eq10}
    V_{inv}(s) = C_{H_{\infty}}(s)[V_{ref}(s)-V_C(s)].
\end{align}
As a result, the closed loop system equation can be found by combining \eqref{eq5} and \eqref{eq10} and can be written as
\begin{align}\label{eq11}
    V_C(s) = G(s)V_{ref}(s) - Z(s)I_d(s),
\end{align}
where
\begin{align*}
    G(s) = \dfrac{G_v^{\Delta}(s)C_{H_{\infty}}(s)}{1+G_v^{\Delta}(s)C_{H_{\infty}}(s)},~~ Z(s) = \dfrac{Z_d^{\Delta}(s)}{1+G_v^{\Delta}(s)C_{H_{\infty}}(s)}. 
\end{align*}
Therefore, it is equivalent to state that the optimal controller is required to be designed such that
\begin{align*}
    &G(s)|_{s=j\omega_o} \approx 1\angle 0^o,\\ 
    &Z(s)|_{s=jh\omega_o} << 1 ~~[h=1, 3, 5, 7, 9, 11, 13].
\end{align*}
The control system of Fig.~\ref{fig:control}\subref{control1} can be realized as a general control configuration as shown in Fig.~\ref{fig:control}\subref{control2}\cite{robust1}. It has a generalized MIMO plant, $P$, containing nominal plant models defined as $G_v^{Nom}(s)$, $Z_d^{Nom}(s)$, $W_S(s)$, $W_{CS}(s)$, $W_d(s)$ and $T_{des}(s)$ with exogenous input signal $w:=\begin{bmatrix}v_{ref} & i_d\end{bmatrix}^\top$ and output signals $z:=\begin{bmatrix}z_S & z_{CS} & e_o\end{bmatrix}^\top$. The controller, $C_{H_{\infty}}(s)$ has input feedback signal, $e$, and output control signal, $v_{inv}$. The uncertainty function, $\Delta(s)$, with input signal, $z_{\Delta}$, and output, $w_{\Delta}$ is represented using upper linear fractional transformations \cite{robust1}. The goal is to synthesize the stabilizing controller, $C_{H_{\infty}}(s)$, that satisfies the following:
\begin{align}\label{eq12}
    ||T_{w\rightarrow z}||_{\infty} < 1.
\end{align}
\begin{remark}
Small-gain theorem states that the accomplishment of \eqref{eq12} together with $||\Delta(s)||_{\infty}\leq 1$ guarantees the robust stability of the closed-loop system.\cite{robust1}
\end{remark}
Using $\gamma$-iteration algorithm provided in \cite{robust1} (the state-space solution approach via solving two algebraic Riccati equations), the synthesis of optimal controller is achieved by means of \textit{hinfsyn} command of MATLAB toolbox. Usually $\mathcal{H}_{\infty}$-control algorithms produce controllers of higher order and model reduction becomes essential to design a low order implementable controller. It is achieved by using balanced truncation method removing modes faster than the switching frequency. The resulting controller, $C_{H_{\infty}}(s)$, is of the order of $17$ which is higher only by $3$ orders than that of conventional PR controller with up to $13^{th}$ order harmonic compensators. Fig.~\ref{fig:bode}\subref{fig:GZmag} and \ref{fig:bode}\subref{fig:GZphase} corroborate the accomplishment of the objectives by the resulting optimal controller. Moreover, the optimal controller along with the plant results $||T_{w\rightarrow z}||_{\infty}=0.77$ that makes the closed-loop system robust stable. Moreover, robust stability margin is verified by \textit{robstab} command of MATLAB. 
\section{Results}
SIMULINK/MATLAB simulation studies are carried out for validation. The system parameters used in both the stages are tabulated in Table~\ref{table}. A combination of impedance type and current source type load models are used. However, for closer proximity with practical scenarios, a DC-drive motor load is also considered along with the others and the modelling is followed as mentioned in \cite{loadmodel1,loadmodel2}.
\renewcommand{\arraystretch}{1.2}
\begin{table}[t]
\centering
\caption{vsi parameters for matlab/simulink and chil simulation}
\label{table}
\begin{tabular}{|c|c|}
\hline 
\textbf{VSI Parameter}              & \textbf{Value}    \\ \hline \hline
Rated RMS Output Voltage               & $220$ (V)         \\ \hline
Rated Output Frequency ($\omega_o$) & $2\pi 60$ (rad/s) \\ \hline
DC Link Voltage ($V_{dc}$)          & $500$ (V)         \\ \hline
Switching Frequency          & $20$ (kHz)         \\ \hline
$LC$ Filter Inductance ($L_f$)      & $2$ (mH)          \\ \hline
$LC$ Filter Capacitance ($C_f$)     & $20$ ($\mu$F)     \\ \hline
Rated Output Power                  & $2$ (kVA)          \\ \hline 
\end{tabular}
\end{table}
\begin{figure*}[t]
	\centering
	\subfloat[]{\includegraphics[scale=0.182,trim={1.0cm 1.0cm 0.0cm 1.0cm},clip]{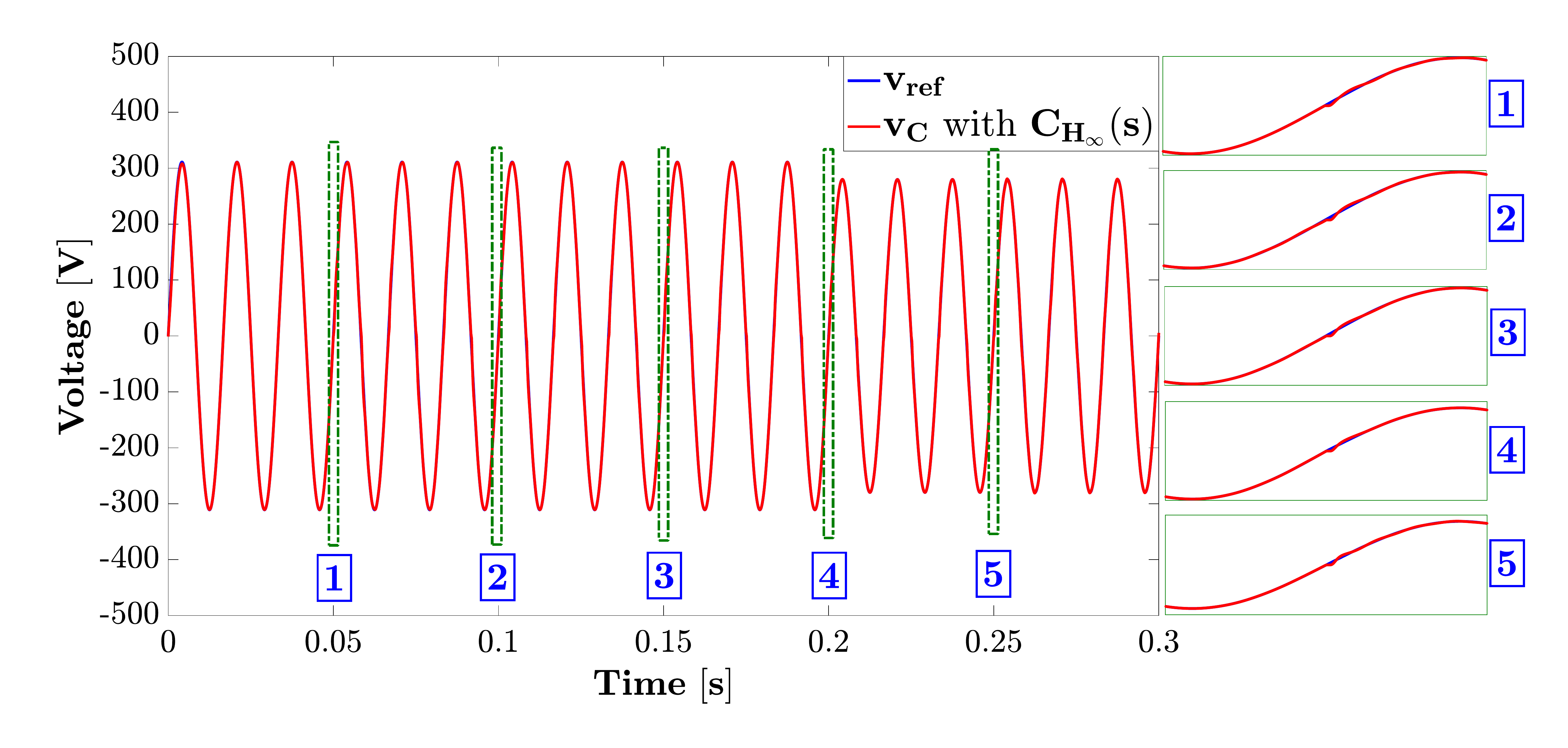}%
	\label{fig:vCHinf}}
	\subfloat[]{\includegraphics[scale=0.182,trim={1.0cm 1.0cm 0.0cm 1.0cm},clip]{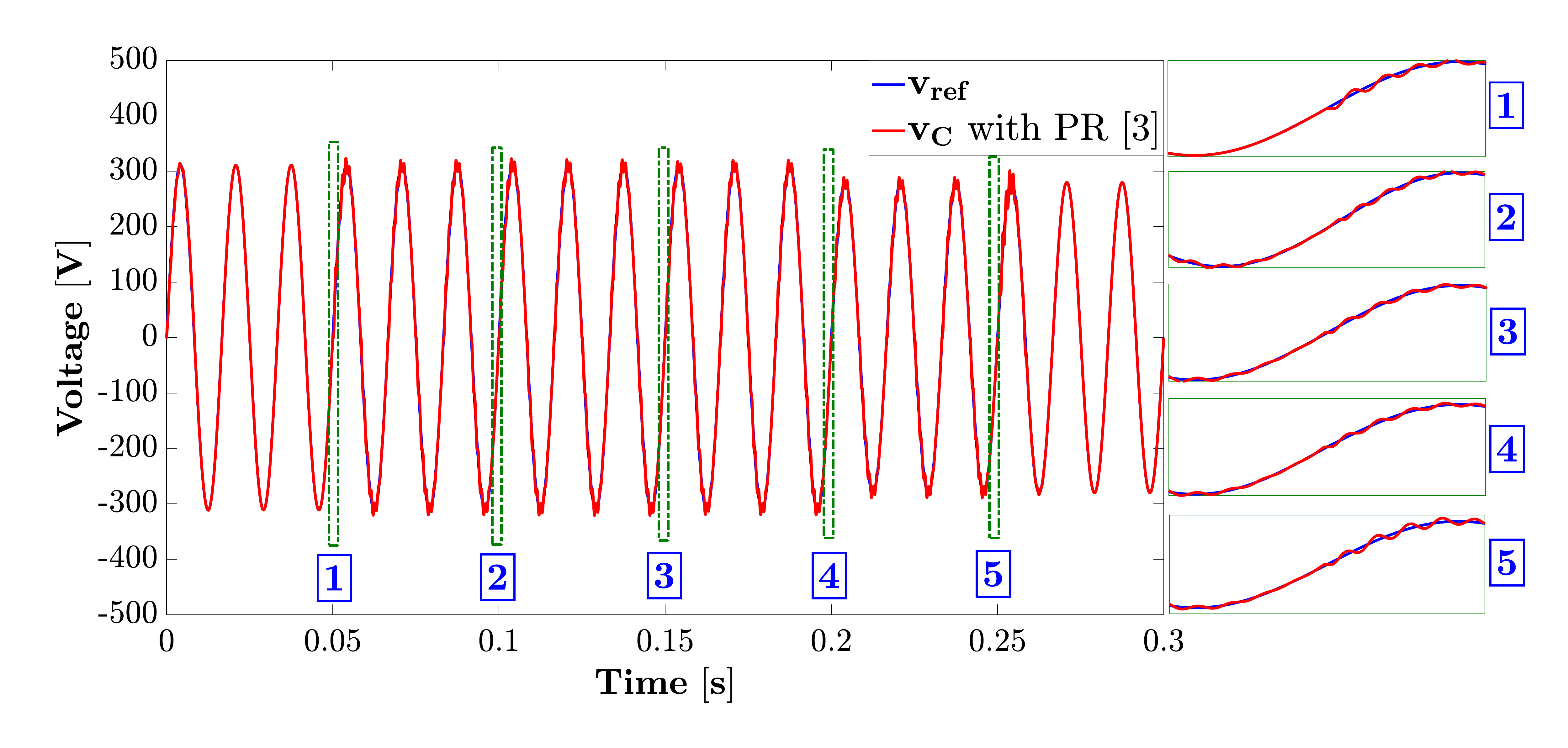}%
	\label{fig:vCPR}}
	\caption{Simulation results, (a) waveform of reference voltage and output voltage with proposed controller, (b) with conventional PR-based controller\cite{yazdani}.}
	\label{fig:sim}
\end{figure*}
\begin{figure}[t]
	\centering
    \includegraphics[scale=0.185,trim={1cm 0.0cm 0.0cm 0.0cm},clip]{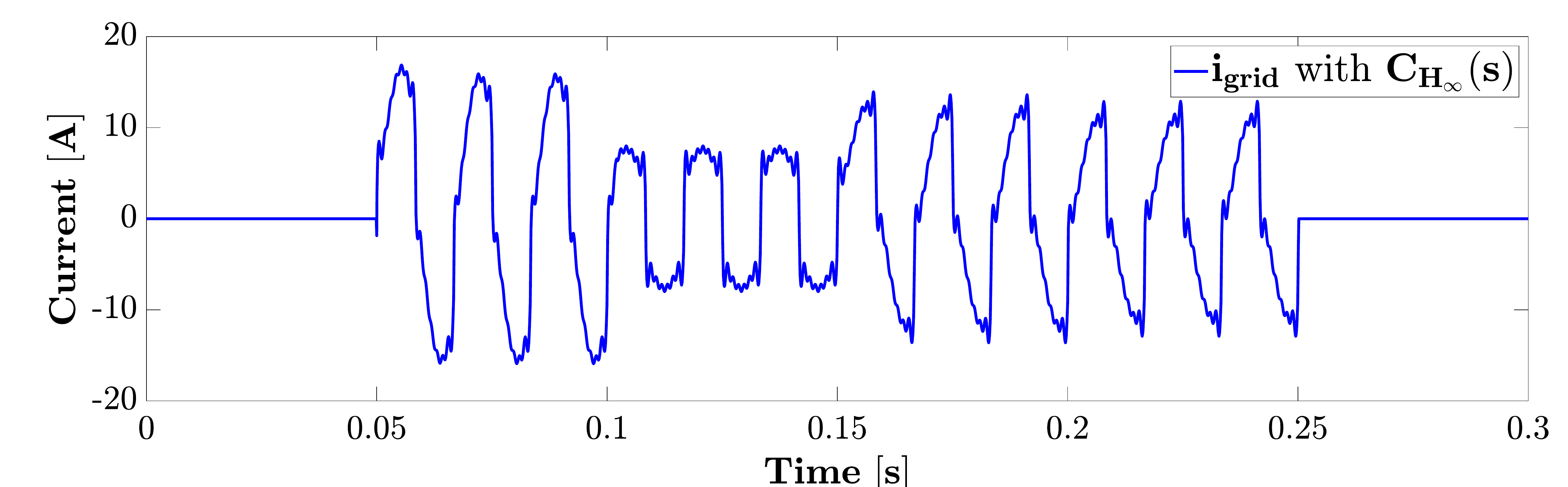}%
	\caption{Simulation results of output load current waveform of VSI.}
	\label{fig:igrid}
\end{figure}
\begin{table}[t]
\centering
\caption{reference tracking and thd performance comparison}
\label{THD}
\begin{tabular}{|c|c|c|c|c|}
\hline 
\textbf{\begin{tabular}[c]{@{}c@{}}Simulation\\ Study\end{tabular}} & \textbf{\begin{tabular}[c]{@{}c@{}}Controller\\ Type\end{tabular}} & \textbf{\begin{tabular}[c]{@{}c@{}}Magnitude \\ Error (\%)\end{tabular}} & \textbf{\begin{tabular}[c]{@{}c@{}}Phase \\ Error (\%)\end{tabular}} & \textbf{\begin{tabular}[c]{@{}c@{}}THD \\ (\%)\end{tabular}} \\ \hline \hline
\multirow{2}{*}{SIMULINK} & $C_{H_{\infty}}(s)$-based & $\pm~0.74$ & $\pm~1.39$ & $0.41$ \\ \cline{2-5} & PR-based\cite{yazdani} & $\pm~1.41$ & $\pm~2.92$ & $1.39$ \\ \hline \hline  CHIL & $C_{H_{\infty}}(s)$-based & $\pm~0.97$ & $\pm~1.89$ & $0.57$\\ \hline  
\end{tabular}
\end{table}
\subsection{Simulation Results and Performance Comparison}
For the purpose of performance comparison, conventional multi-loop-based classical outer PR-based voltage controller and inner PI-based current controller is considered for the time-domain simulation. \cite{yazdani} provides an elaborated guidelines for designing the classical PR and PI based multi-loop voltage controller for grid-forming VSI with enough gain and phase margins. In this work, the guidelines are followed in order to possess a fair comparative study. PI-based inner loop is designed to have phase margin (PM) $\geq 60^{\circ}$ and gain margin (GM) $\geq 40$~dB with bandwidth $\geq 5$~kHz. Similarly, PR-based outer loop is designed to have PM $\geq 45^{\circ}$ and GM $ \geq 40$~dB with bandwidth $\geq 5$~kHz. The following sequence of events is used in the MATLAB/SIMULINK-based validation stage.
\begin{itemize}
    \item VSI is in no-load condition until $t=0.05~s$. 
    \item At $t=0.05~s$, there is a transition from no-load to full-load and stays at full-load condition until $t=0.1~s$.
    \item At $t=0.1~s$, reactive power drops to $0$ until $t=0.15~s$.
    \item At $t=0.15~s$, both real and reactive power change.
    \item At $t=0.2~s$, $v_{ref}$ drops and stays until $t=0.25~s$.
    \item At $t=0.25~s$, VSI is switched to no-load condition. 
\end{itemize}
The output voltage waveform of VSI, $v_C$, during the time-domain events with both proposed $\mathcal{H}_{\infty}$-based and conventional-multi-loop-based voltage controller are shown in Fig.~\ref{fig:sim}\subref{fig:vCHinf} and Fig.~\ref{fig:sim}\subref{fig:vCPR}. The enlarged portions in these figures are illustrating the transient behaviors during aforementioned situations. The results substantiate the fact that the proposed $\mathcal{H}_{\infty}$-based controller exhibits superior robustness in performance than the multi-loop controller with varying loading conditions. THD of $v_C$ with proposed controller is much smaller than the multi-loop controller with highly non-linear loads (output current is illustrated in Fig.~\ref{fig:igrid}). Finally, Table \ref{THD} corroborates the superiority of the proposed controller in reference tracking and disturbance rejection in compare with the classical multi-loop PR-based voltage controller at the cost of increasing the order of controller only by $3$. 
\begin{figure}[t]
	\centering
    \includegraphics[scale=0.33,trim={0.0cm 0.0cm 27cm 0.0cm},clip]{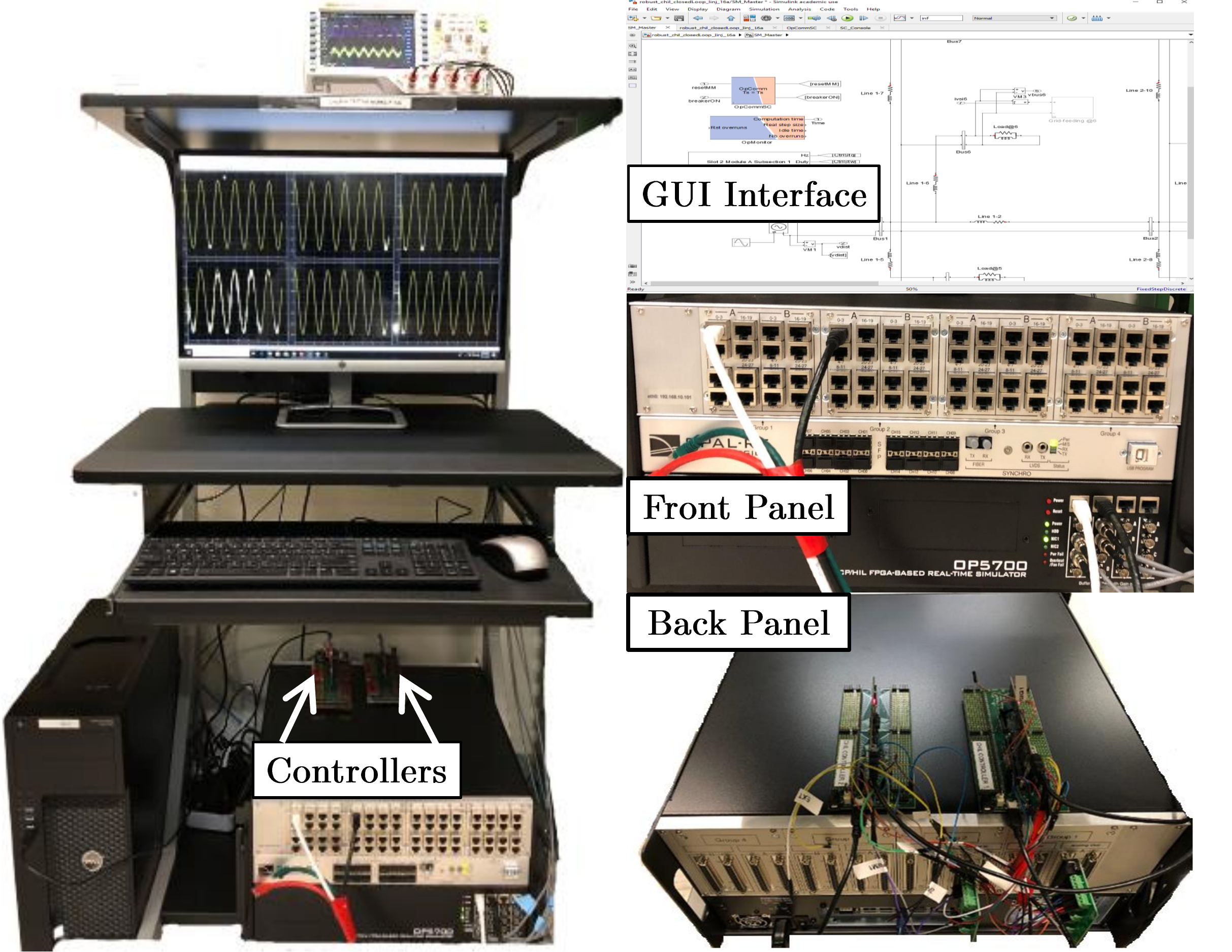}%
	\caption{OPAL-RT based hardware-in-the-loop simulation platform with Texas Instruments Delfino TMS320F28379D controller board.}
	\label{fig:chilplatform}
\end{figure}
\subsection{CHIL Simulation Results}
The computational footprint of the proposed $\mathcal{H}_{\infty}$-based controller, while implemented in a real low-cost micro-controller board, is an essential check for further validating the performance. Therefore, controller hardware-in-the-loop (CHIL)-based simulation studies are also conducted on OPAL-RT real-time simulator with the proposed $\mathcal{H}_{\infty}$-based controller realized on a low-cost Texas-Instruments TMS28379D Delfino controller board as shown in Fig.~\ref{fig:chilplatform}. In similar way, four events are emulated in the OPAL-RT platform and are enlisted as:
\begin{itemize}
    \item $\mathtt{CASE}$-$\mathtt{1}$: A sudden rise in VSI output current, $i_{grid}(t)$, at $t=158.05~s$ with unchanged reference signal, $v_{ref}(t)$.
    \item $\mathtt{CASE}$-$\mathtt{2}$: A sudden fall in VSI output current, $i_{grid}(t)$, at $t=130.35~s$ with unchanged reference signal, $v_{ref}(t)$.
    \item $\mathtt{CASE}$-$\mathtt{3}$: A sudden fall in output voltage reference signal, $v_{ref}(t)$, at $t=132.95~s$ with unchanged $i_{grid}(t)$.
    \item $\mathtt{CASE}$-$\mathtt{4}$: A sudden rise in output voltage reference signal, $v_{ref}(t)$, at $t=166.05~s$ with unchanged $i_{grid}(t)$.
\end{itemize}
\begin{figure*}[t]
	\centering
	\subfloat[]{\includegraphics[scale=0.19,trim={0.0cm 6.5cm 5.0cm 1.0cm},clip]{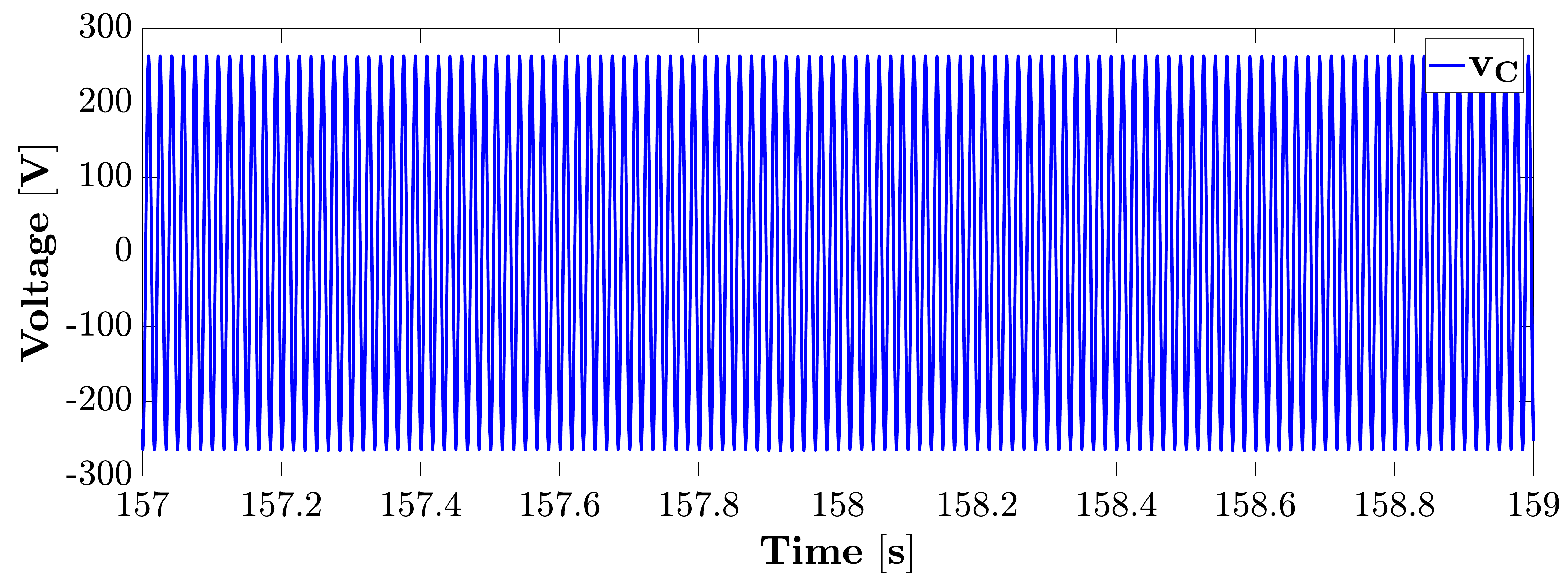}%
	\label{fig:vCHinfchil1}}
	\subfloat[]{\includegraphics[scale=0.19,trim={0.0cm 6.5cm 5.0cm 1.0cm},clip]{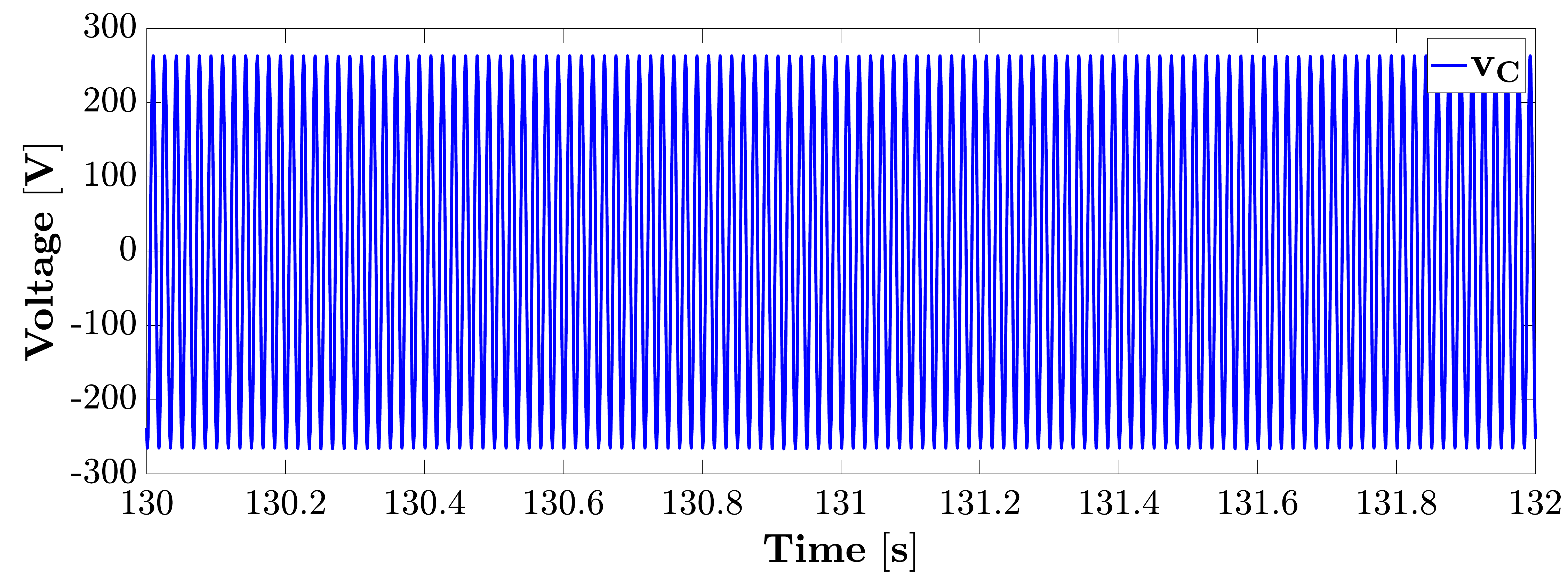}%
	\label{fig:vCHinfchil2}}
		
    \subfloat[]{\includegraphics[scale=0.19,trim={0.0cm 6.5cm 5.0cm 1.0cm},clip]{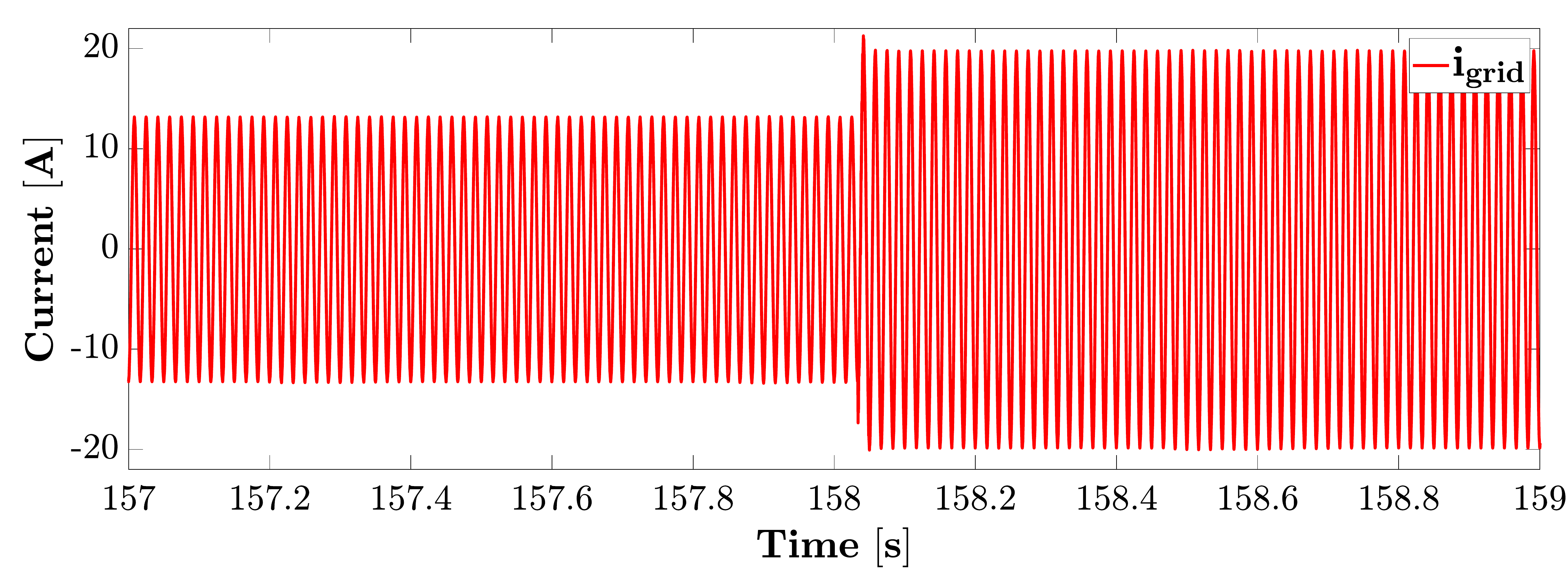}%
	\label{fig:igridchil1}}
	\subfloat[]{\includegraphics[scale=0.19,trim={0.0cm 6.5cm 5.0cm 1.0cm},clip]{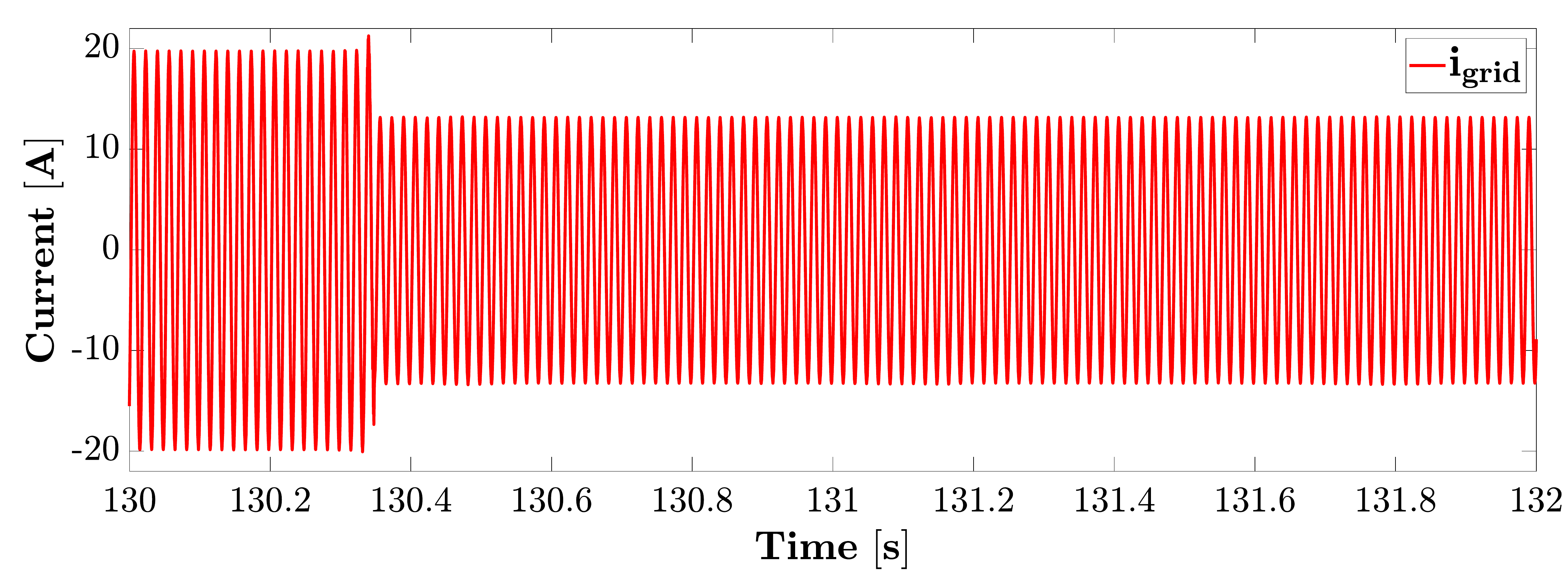}%
	\label{fig:igridchil2}}
	\caption{CHIL simulation results of output voltage waveform of VSI with (a) $\mathtt{CASE}$-$\mathtt{1}$: sudden rise in output current at $t=158.05~s$ and (b) $\mathtt{CASE}$-$\mathtt{2}$: sudden fall in output current at $t=130.35~s$ with proposed controller, (c) rise in output current in $\mathtt{CASE}$-$\mathtt{1}$, (d) fall in output current in $\mathtt{CASE}$-$\mathtt{2}$.}
	\label{fig:chil1}
\end{figure*}
\begin{figure*}[t]
	\centering
	\subfloat[]{\includegraphics[scale=0.19,trim={0.0cm 6.5cm 5.0cm 1.0cm},clip]{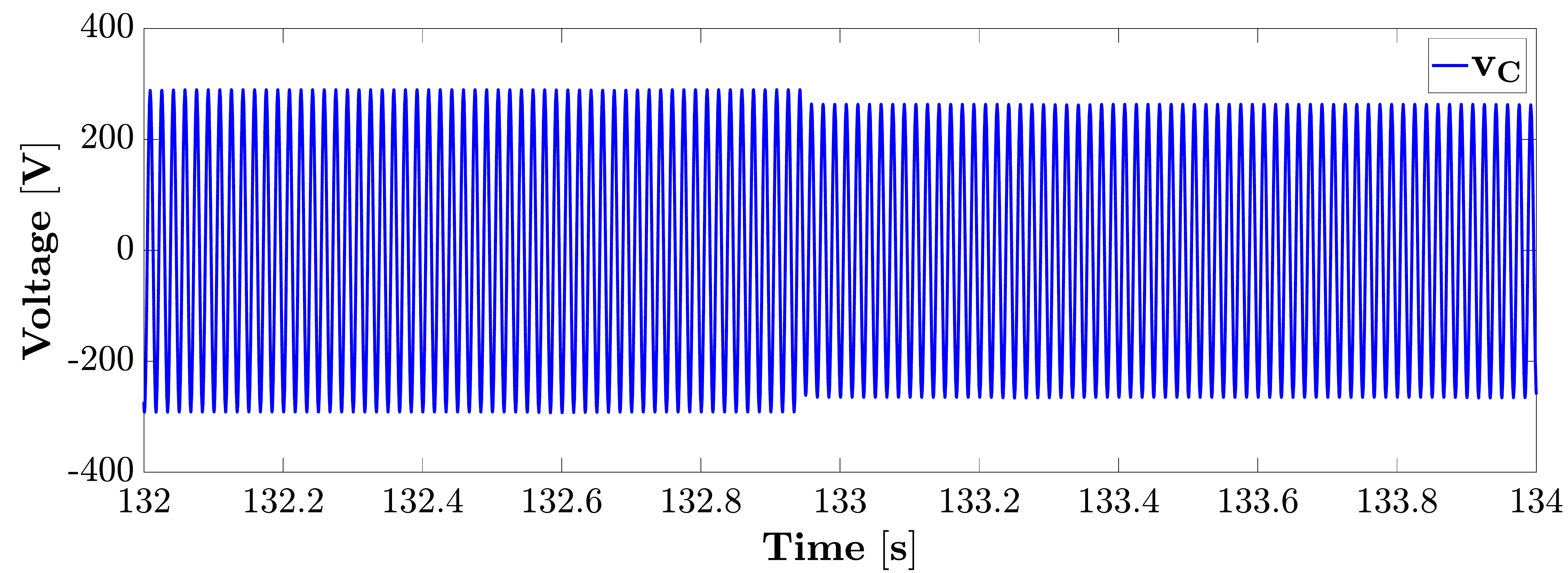}%
	\label{fig:vCHinfchil3}}
	\subfloat[]{\includegraphics[scale=0.19,trim={0.0cm 6.5cm 5.0cm 1.0cm},clip]{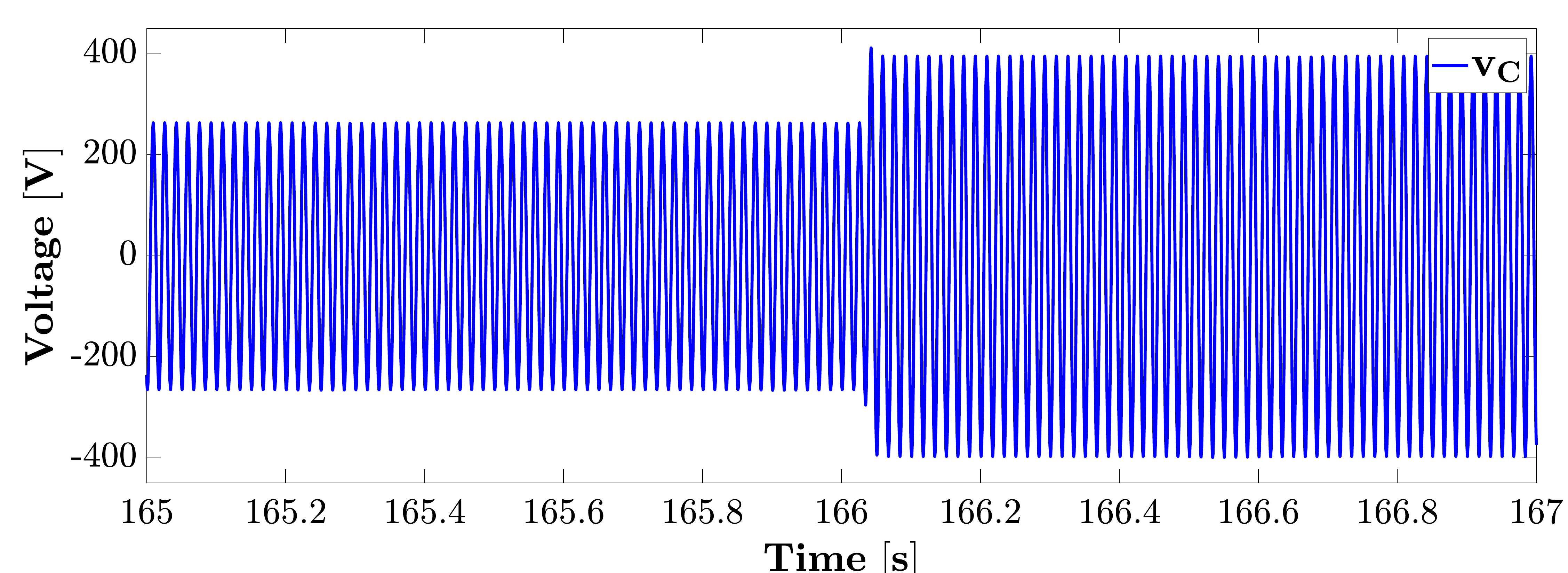}%
	\label{fig:vCHinfchil4}}
		
    \subfloat[]{\includegraphics[scale=0.19,trim={0.0cm 6.5cm 5.0cm 1.0cm},clip]{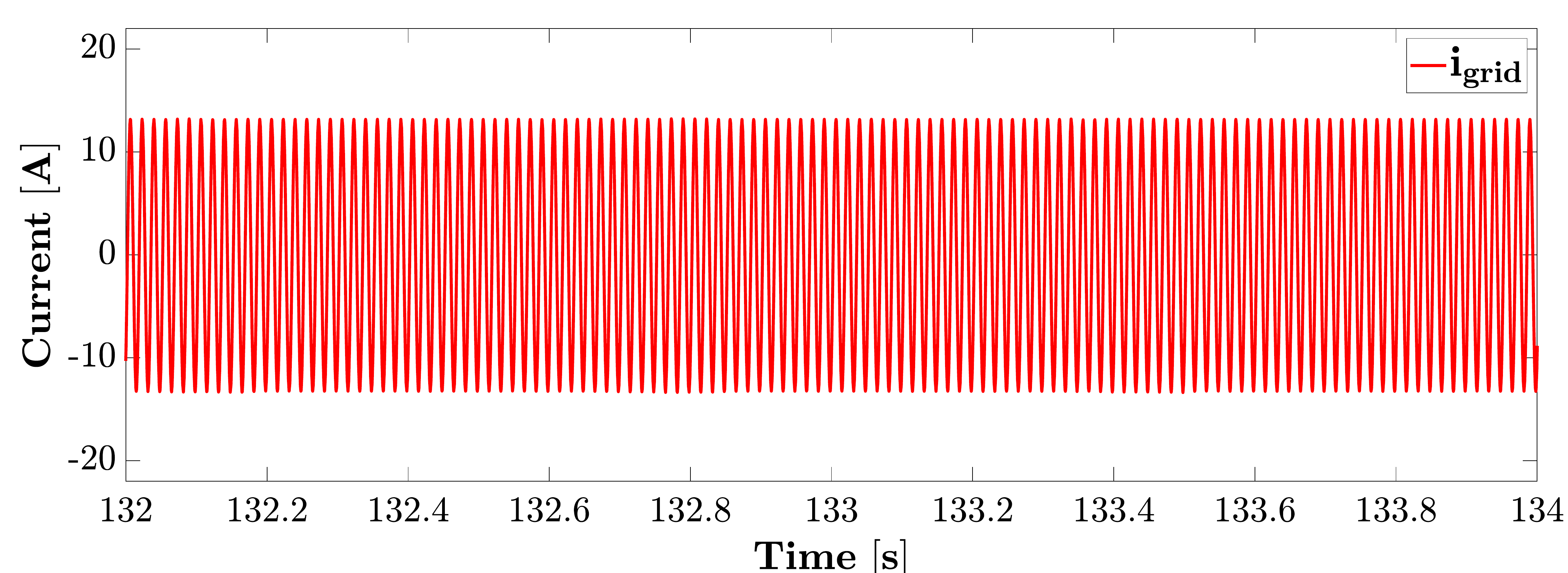}%
	\label{fig:igridchil3}}
	\subfloat[]{\includegraphics[scale=0.19,trim={0.0cm 6.5cm 5.0cm 1.0cm},clip]{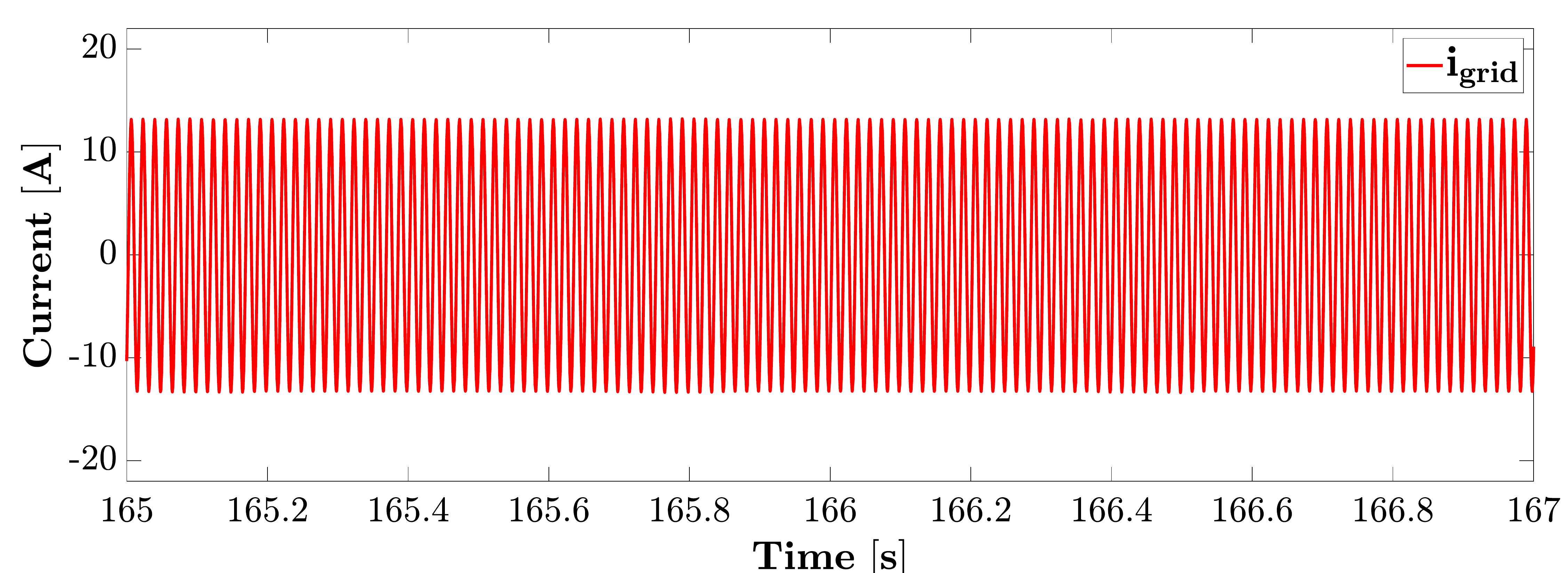}%
	\label{fig:igridchil4}}
	\caption{CHIL simulation results of output voltage waveform of VSI with (a) $\mathtt{CASE}$-$\mathtt{3}$: sudden fall in output voltage reference signal at $t=132.95~s$ and (b) $\mathtt{CASE}$-$\mathtt{4}$: sudden rise in output voltage reference signal at $t=166.05~s$ with proposed controller, (c) unchanged output current in $\mathtt{CASE}$-$\mathtt{3}$ and (d) in $\mathtt{CASE}$-$\mathtt{4}$.}
	\label{fig:chil2}
\end{figure*}
The results for $\mathtt{CASE}$-$\mathtt{1}$ and $\mathtt{CASE}$-$\mathtt{2}$ are shown in Fig.~\ref{fig:chil1}. It is clearly observed that the output voltage waveforms of VSI, $v_C$, are insensitive to the sudden changes of $i_{grid}$ (shown in Fig.~\ref{fig:chil1}\subref{fig:igridchil1} and Fig.~\ref{fig:chil1}\subref{fig:igridchil2}) in both $\mathtt{CASE}$-$\mathtt{1}$ and $\mathtt{CASE}$-$\mathtt{2}$ as shown in Fig.~\ref{fig:chil1}\subref{fig:vCHinfchil1} and Fig.~\ref{fig:chil1}\subref{fig:vCHinfchil2}. Similarly, the results for $\mathtt{CASE}$-$\mathtt{3}$ and $\mathtt{CASE}$-$\mathtt{4}$ are shown in Fig.~\ref{fig:chil2}. It is observed that the output voltage waveforms of VSI, $v_C$, tracks the changes in reference voltage signal, $v_{ref}$, very quickly as shown in Fig.~\ref{fig:chil2}\subref{fig:vCHinfchil3} and Fig.~\ref{fig:chil2}\subref{fig:vCHinfchil4} for both $\mathtt{CASE}$-$\mathtt{3}$ and $\mathtt{CASE}$-$\mathtt{4}$. Moreover, due to constant-current type output loads terminated across the VSI, the output current waveform, $i_{grid}$ is unchanged as shown in Fig.~\ref{fig:chil2}\subref{fig:igridchil3} and Fig.~\ref{fig:chil2}\subref{fig:igridchil4}. The results shown in Fig.~\ref{fig:chil1} and Fig.~\ref{fig:chil2} clearly substantiate the fact that the dynamic performance of the proposed controller is less sensitive to transients. Moreover, Table~\ref{THD} shows that the reference tracking and harmonic rejection performance illustrated in CHIL results are close enough to the results from SIMULINK/MATLAB. However, due to practical and inevitable limitations like quantization error of ADC of the real controller and efficacy of discretization process, the results may differ slightly but under acceptable limit as evidenced in this case also. Along with the performances, this study also showcases the viability of the proposed $\mathcal{H}_{\infty}$-based controller in real low-cost control broads.  
\section{Conclusion}
This article demonstrates the design and implementation of robust and optimal single-loop voltage controller for single-phase grid-forming VSI. The model uncertainty of VSI imposed by the unknown changing load is demonstrated and its impact on dynamic model of VSI is shown. $\mathcal{H}_{\infty}$-based controller design is introduced to address this issue and the required objectives for the optimal controller are discussed to formulate the optimization problem which leads to final optimal controller. A time-domain SIMULINK-based simulation study substantiates the fact that the resulting controller exhibits superior robustness in performance during varying loading of VSI that conventional multi-loop controller architecture. Moreover, OPAL-RT based CHIL simulations are conducted to verify the viability of the resulting controller.
\section*{Acknowledgment}
The authors acknowledge Advanced Research Projects Agency-Energy (ARPA-E) for supporting this research through the project titled ``Rapidly Viable Sustained Grid'' via grant no. DE-AR0001016.
\bibliographystyle{IEEEtran}
\bibliography{reference}
\end{document}